# Innovative in silico approaches to address avian flu using grid technology

Vincent BRETON[a,*], Ana Lucia DA COSTA[a], Paul DE VLIEGER[a], Lydia MAIGNE[a], David SARRAMIA[a], Young-Min KIM[b], Doman KIM[b], Hon Quang NGUYEN[c], Tony SOLOMONIDES[d], Yin-Ta WU[e], Truong Nam HAI[f]

[a] *Laboratoire de Physique Corpusculaire de Clermont-Ferrand, CNRS-IN2P3, Aubière, France*
[b] *Chonnam National University, Gwangju, South Korea*
[c] *IFI (Institut de la Francophonie pour l'Informatique), Hanoi, Vietnam*
[d] *University of West England, Bristol, United Kingdom*
[e] *Academia Sinica, Taipei, Taiwan*
[f] *Vietnam Academy of Sciences and Technology, Hanoi, Vietnam*

**Abstract**: The recent years have seen the emergence of diseases which have spread very quickly all around the world either through human travels like SARS or animal migration like avian flu. Among the biggest challenges raised by infectious emerging diseases, one is related to the constant mutation of the viruses which turns them into continuously moving targets for drug and vaccine discovery. Another challenge is related to the early detection and surveillance of the diseases as new cases can appear just anywhere due to the globalization of exchanges and the circulation of people and animals around the earth, as recently demonstrated by the avian flu epidemics. For 3 years now, a collaboration of teams in Europe and Asia has been exploring some innovative *in silico* approaches to better tackle avian flu taking advantage of the very large computing resources available on international grid infrastructures. Grids were used to study the impact of mutations on the effectiveness of existing drugs against H5N1 and to find potentially new leads active on mutated strains. Grids allow also the integration of distributed data in a completely secured way. The paper presents how we are currently exploring how to integrate the existing data sources towards a global surveillance network for molecular epidemiology.

## INTRODUCTION

Emerging diseases know no frontier. Public health authorities have identified the risk they represent and developed national and international policies to address them. The challenge to achieve early response is to collect critical data and to quickly process them in order to launch a prompt alert. However, the relevant data are often coming from multiple sources which are widely distributed geographically. Indeed, experience with SARS and Avian Flu has clearly shown that the emerging diseases have to be monitored at a world level. The emergence of these diseases has also triggered the development of epidemiological surveillance networks. These networks are collecting data but do not provide the facilities to process them. International agencies like FAO or WHO make a significant corpus of data publicly available on central servers. These servers provide very useful information to the decision makers and to the public opinion but can not be considered for raising an alert in case of pandemics. More relevant are the specialized molecular biology databases which have been set up for influenza viruses but they do not provide the latest strains sequenced.

Grid technology has been quickly evolving in the last 10 years. Today, it has reached a level of maturity in the field of distributed computing and data management, which makes it a natural choice to implement a surveillance network for emerging diseases. It allows federating distributed data sources, querying them remotely and on demand mobilizing large CPU resources to analyse them. In this paper, we discuss opportunities provided by the grid technology to improve the global response to emerging diseases by improving the collection, availability and accessibility of disease related data for molecular epidemiology and *in silico* drug discovery as well as developing an alert system. The method is to develop an international network of data repositories integrated into a data grid. Through the access and query of these data, services are offered to the research community, allowing the network to work as an alert system for public health authorities. The grid added value compared to the other existing efforts is that no movement of data is needed as they are queried where they are produced. Moreover, services are provided on demand for data analysis on the grid computing resources.

* Address correspondence to this author at the Corpuscular Physics Laboratory of Clermont-Ferrand, CNRS-IN2P3, PCSV Team, 24 avenue des Landais, 63177 AUBIERE, France; Tel: +33 473 407 219; E-mail: breton@clermont.in2p3.fr

Such a network is well fitted to address the two main concerns of the international community which are the evolution of the virus and its capacity for human-to-human transmission, through the set-up of two families of services:
- Molecular epidemiology: once a new virus stream has been sequenced, its comparison to the previously identified streams allows measuring the virus evolution.
- Virtual screening: *in silico* docking allows monitoring the impact of mutations on the existing drugs and rapid identification of new hits on mutated strains.

Today, thank to the progresses in web and grid technologies, building such a surveillance network is more a political than a technical challenge. In this paper, we will provide examples of services currently operated on grid infrastructures, which fit exactly the requirements of the proposed network.

The paper is organized as follows:
- In section 1, we discuss the context of avian flu epidemics on one side and grid technology on the other side.
- In section 2, we present how grids are already used to do virtual screening on neuraminidase N1 and to study the impact of mutations on the efficiency of the existing drugs targeting this enzyme.
- In section 3, we provide an example of a grid enabled disease specific surveillance network in the Auvergne region in France federating databases with patient anatomical data and in a later stage integrating oncogenetics data
- In section 4, we present the vision for a surveillance network on avian flu
- And section 5 presents some conclusions and perspectives.

## CONTEXT

### Medical context: emerging diseases, a growing threat to public health

Avian influenza virus, which has claimed its first human victim in 1997 in Hong Kong, became the primary health concern of the 21st century. Although various strains of avian influenza have been recognized for decades, the « highly pathogenic avian influenza virus of type A of subtype H5N1 » is a likely source of the next human influenza pandemic **[1].** In its latest report from September 2008, WHO reports 387 human cases in 15 countries and 245 deaths between 2003 and 2008, mostly in Asia **[2]**. H5N1 is primarily a pathogen of poultry; more than 200 million poultry died because of this virus. Human acquired the virus by having direct or close contact with H5N1-infected poultry or poultry products **[3]**. More than 60 % of human infections have been fatal, so far the spread of avian influenza viruses from one ill person to another has been reported very rarely. But the biggest fear remains the risk of the reassortment of an avian influenza virus with a human (or other non-avian) influenza virus, more dangerous and difficult to combat **[4]**. The ability of the international community to respond efficiently to the possible emergence of a human-to-human transmissible avian influenza virus depends on its capacity to quickly assess any evolution of the disease. Since 1997, the virus has spread through Asia, the Middle East, Europe and Africa. This global and rapid spread of H5N1 induces the necessity to improve pandemic preparedness and communication, early detection and surveillance, containment and responses systems. Use of drugs against influenza virus could represent a first line defence against a new pandemic, allowing the control of the infection until sufficient quantities of a suitable vaccine can be produced. Zanamivir (Relenza®) and oseltamivir (Tamiflu®) are available antiviral molecules inhibiting a key viral protein: neuraminidase (NA). Unfortunately, the report of development of drug resistance causes growing concern **[5].**

Grid technology, presented in next section, is a relevant choice to address evolution and transmission of avian influenza by setting up a surveillance network for federating distributed data, querying them remotely and performing analyses on demand**.**

**Technical context: grids, an emerging technology for mutualising and federating data and computing resources**

Grid computing is an exciting new technology promising to revolutionise many services already offered by the internet. Grids are defined as a fully distributed, dynamically reconfigurable, scalable and autonomous infrastructure to provide location independent, pervasive, reliable, secure and efficient access to a coordinated set of services encapsulating and virtualizing resource **[6]**. This new paradigm offers rapid computation, large-scale data storage and flexible collaboration by harnessing together the power of a large number of commodity computers or clusters of other basic machines. The grid was devised for use in scientific fields, such as particle physics and bioinformatics, in which large volumes of data, or very rapid processing, or both, are necessary. Unsurprisingly, the grid has also been used in a number of ambitious medical and healthcare applications such as MammoGrid **[7]** or Health-e-Child **[8]**. A 'healthgrid' is an innovative use of this emerging information technology to support broad access to rapid, cost-effective and high quality healthcare **[9]**.

Well-identified areas of relevance of the grid paradigm are epidemiology and drug discovery. Epidemiology focused on population-level research requires access to distributed, critically sensitive and heterogeneous data, resulting in overall costly computing processes. Users ought to be able to take for granted that the security mechanisms are sufficient to protect their data, that the results of their research will be private and available to third parties only if desirable, that the system will meet the concerns of the ethical and legal committees of their research institutions, that the services are reliable, efficient and permanent, that they do not have to change significantly their current procedures, protocols or workflow and finally that the data is somehow automatically organised and gathered, thus available for further exploitation. Early attempts at epidemiological applications of grids **[10]** have demonstrated their relevance for patient customized research.

The discovery and development of new drugs is very costly and attrition rates are high. Initiatives to reduce the rate of attrition during later phases of development are clearly desirable and if successfully implemented will reduce costs. Predictive pharmacology at the discovery research stage has been identified as one of the key bottlenecks in the pharmaceutical R&D process **[11]**. As well, knowledge management is essential to leveraging the potential of genomics and proteomics and to analyse the huge amount of information in an integrated way. In this context, the computing power of the grids is starting to be exploited for the identification of new promising compounds through virtual screening **[12]**.

In the next section, we are going to present briefly some significant results obtained in the field of drug discovery on avian flu thank to grid infrastructures.

**CURRENT USES OF GRID TECHNOLOGY FOR DRUG DISCOVERY ON AVIAN FLU**

**In silico virtual screening**

In 2005, the WISDOM (Wide In Silico Docking On Malaria) addressed successfully *in silico* high-throughput screening on grid infrastructures to look for new inhibitors against enzymes implicated in malaria, reducing the time and the cost of pharmaceutical development for this neglected disease **[13]**. Based on this first successful experience, in 2006, it was decided to perform a computational challenge to deploy large virtual screening of 300,000 chemical compounds selected from ZINC database (http://zinc.docking.org/) and a chemical combinatorial library against 8 variants of neuraminidases predicted by homology modelling. NA structures were predicted because no NA subtype 1 (NA1) was available for structural study. Afterwards, crystal structures published for NA1 **[14]** were in good agreement with our modelled structures with a RMSD of 1.4 Å (all atoms). In order to model complexes of candidate compounds to structures of NA binding sites, the Autodock program [15] was deployed. This program allows fast calculation of binding energies of possible binding poses. Compounds are ranked based on docking score and binding mode. Grid technology enables to provide the computational and storage resources required for the *in silico* experiment. During six weeks, 2000 CPUs on the EGEE (Enabling Grid for E-sciencE) grid infrastructure have been mobilized [16] to perform 4 million docking calculations, which would require 100 years on one single computer. Two grid environments were used in this data challenge in order to have efficient and interactive control of massive molecular dockings: WISDOM and DIANE **[17]**. About 600 gigabytes of output data were produced and needed to be post-processed to avoid false positives. Few hundred chemical compounds were finally selected and tested in a biochemical laboratory in order to evaluate in vitro inhibition activity.

A wild type NA from H5N1 influenza virus strain A/Vietnam/1203/04 was prepared using *E. coli* expression systems and purified to single protein on a SDS-PAGE. NA activity was determined using MUNANA [2'-(4-methylumberlliferyl)-alpha-D-N-acetyl neuraminic acid] as a fluorogenic substrate **[18]**. Inhibition activity of NA was determined by incubating enzyme solution with 40 mM sodium phosphate buffer (pH 7.2), MUNANA, and with or without ligands. Fluorescence was read with a fluorescence plate reader SoftMax Pro 5 (Molecular Devices, USA) using excitation and emission wavelengths of 362 nm and 448 nm. Among 185 compounds tested 79 (42.9%) compounds showed inhibition for NA activity of H5N1, and 59 compounds (31.9%) showed same or better activity than that of oseltamivir which is a general inhibitor for NA of H5N1. In near future, mutant forms of NA will be tested in laboratory to allow correlation with results obtained *in silico* explained in next paragraph.

Successfully validated *in vitro*, this virtual screening process can be routinely performed when a new NA mutation appears allowing a fast *in silico* answer which can be really useful to evaluate the impact of enzyme changes. Moreover, other key viral proteins can be investigated. Thanks to the grid large storage capacity and smart data management services, virtual data produced can be massively stored and easily accessed.

**Study of the impact of mutations on the efficiency of existing drugs**

NA1 structures, including one with original sequence (noted as original type) and five with different single point mutation, namely T01 (E119A), T02 (E119D), T03 (H274F), T04 (H274Y), and T05 (R293K), in their binding sites, were established. Additionally, based on molecular dynamics simulation, a unique amino acid, Y344, in NA1 can pose either open or close conformation to binding pocket. The open conformer is included in original type, noted T13, and E119A mutant, noted T07, for measuring its effects.

A library of 308,585 compounds, including five known active compounds (Scheme. (**1**)), was docked into the binding site of NA1 using Autodock3.0.5 [**15**]. The docking energy of each pose was computed by Autodock and was used as an indication of inhibiting potential. The known active compounds were applied as the threshold of screening hits.

As shown in Fig. (**1**), after ranking docked poses by their computed docking energies, the compound library distributed normally among -18kcal/mol to -2 kcal/mol. Two known potent drugs (GNA:zanamivir; G39: oseltamivir) presented within the top 5% of all compounds against target T06 (original type); two less potent inhibitors (DAN and 4AM) fell behind top 15%, which difference should be enough to discriminate between hits and non-hits. Furthermore, by referring to both positive controls, large portion of less active compounds (5% cutoff) were successfully filtered. In overall, our modelling effectively extracts potential hits and results in a suggested focus library for experimental assay.

Changes of compound docking energies occurred in most mutants (Fig. (**2**)), the variants T01(E119A), T03(H274F), and T05(R293K) have greater impacts on the hits, whose docking energy better than -14.0 kcal/mol against T06. This observation agreed with other studies reported in the literature [**19, 20**]. It was also interesting to notice that the "open" and "close" conformation of Tyr344 (in T07 and T13) has obvious effects on the compound binding. Therefore, incorporation of interacting with Y344 should be considered in the future drug design. Knowing this effect has resulted in synthesis of highly potent inhibitor [**21**].

Mutation effects of known inhibitors may be predicted with similar method, as illustrated in Fig. (**3**), the docking energy of GNA (zanamivir) dropped from -14.7 kcal/mol to -12.7kcal/mol owing to the mutation of Glu119 to Ala (target T01). In general, G39 (oseltamivir) was more sensitive to the mutation than GNA at the sites of E119 and H274 (Table. (**1**)).

The next section introduces the use of distributed data for cancer surveillance network and the benefits of a grid environment in this case for secured access on data.

**GRIDS FOR EPIDEMIOLOGY: EXAMPLE OF A CANCER SURVEILLANCE NETWORK**

We would like to illustrate the interest of the grid approach for accessing distributed biological or medical data by taking the example of a surveillance network currently under design in the Auvergne region in the field of oncology. This network aims at improving the exchange of information between the public and private actors involved in the surveillance of cancers including the anatomical pathology, cytopathology laboratories, the associations in charge of the early detection programs, the local and national epidemiological services and possibly the patients. In France, the national program for early diagnosis of breast cancer is operated by associations, which are in charge of inviting through advertisement all women above 50 years old to undergo breast examination by mammography every three years. If some women are positively detected with tumours, the associations are in charge of providing a second diagnosis on the mammograms and following up on the medical process the patients are undergoing. This involves collecting the anatomical pathology data about the tumour which are stored by the laboratories in charge of biopsies. Presently, the patient data stored in laboratories are faxed on request or carried out by hand to the associations where they are registered again on a local computer. This process is costly and errors prone as data have to be typed and reinterpreted twice. Another alternative is for the associations to query the databases of the anatomical pathology laboratories on demand. The grid technology is particularly well fitted for such an approach: indeed, a grid federating the databases (see Fig. (**4**)) would provide a secured framework where the patient data are left in the laboratories and where some identified users from the associations deploy distributed queries on the data to fill their local patient file. Other customers of the data in the anatomical pathology laboratories are the epidemiological services of the National Institute for Sanitary Watch (Institut National de Veille Sanitaire) and of the regional epidemiological observatory. These two structures are just interested by a set of statistical indicators without any knowledge of the patient name.

In the surveillance network we are proposing, the use of grid technology allows leaving all the data in the laboratories where they are produced. Moreover, the grid security framework allows also granting to the anatomical pathology laboratories a complete control on the access rights to their data. This should reduce the fear to loose control on their data. The network targets initially the actors of cancer surveillance but the patients could be considered as users in a second step as shown in Fig. (**4**). The use case is presently being implemented in the French Auvergne region using grid technology developed by the EGEE and Auvergrid projects.

In the next section, we describe the architecture and services of a surveillance network on avian flu built using grid technology, which could provide useful services to the research community for both drug discovery and epidemiology.

**Scheme (1).** Structures of the five known inhibitors in the library.

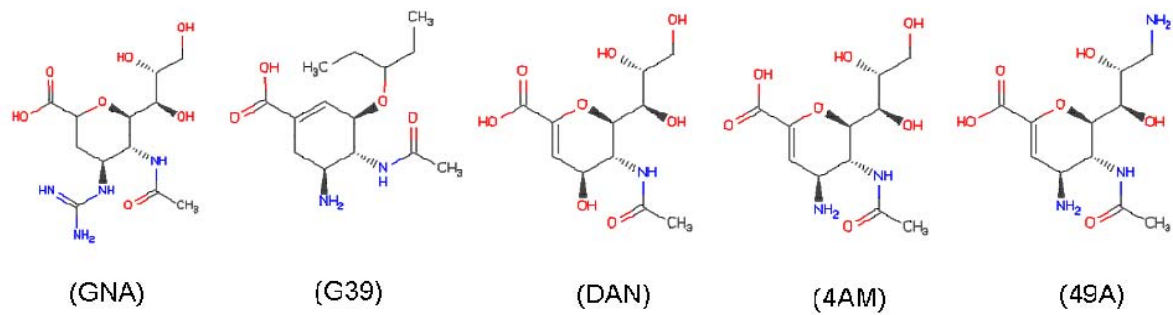

**Fig (1).** Histogram of compounds against T06. Those portions that control compounds fall into are colored in red.

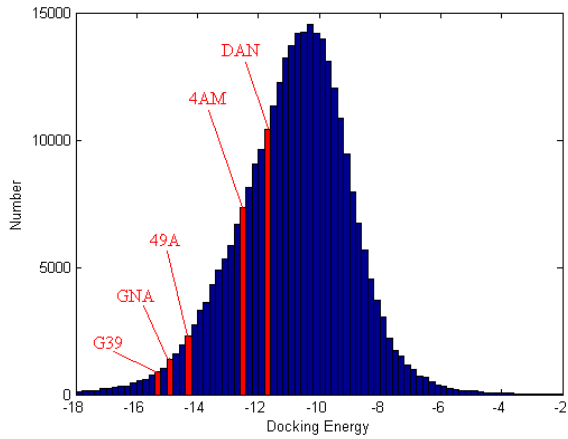

**Fig (3).** Histogram of compounds against T01. Control compounds are shown to decrease docking energies in this variant.

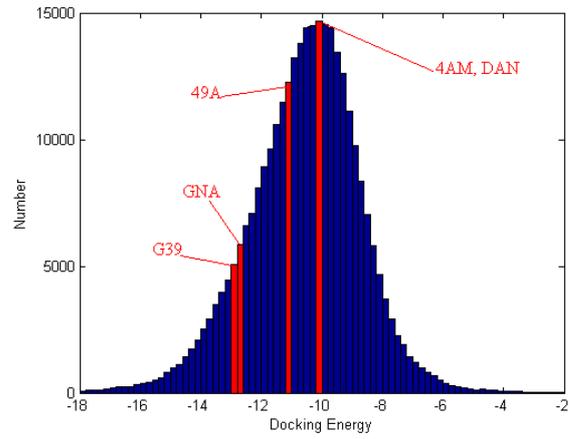

**Fig (2).** Heat map of compounds ordered by their docking energies against T06 and colored by their docking energies against each target. It is clear to see that single site mutation affect compound binding to targets. The effect is more obvious in mutation targets T01, T03, and T05.

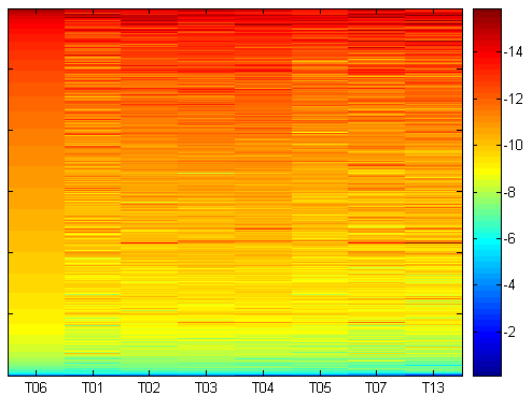

**Table (1).** Summary of mutation effect on the control compounds from the modelling. The "+" sign indicates the compound is within top 5%; The "-" sign indicates the compound is not within the top 5%.

|     | T06 | T01 | T02 | T03 | T05 | T07 | T13 |
|-----|-----|-----|-----|-----|-----|-----|-----|
| DAN | -   | -   | -   | -   | -   | -   | -   |
| 4AM | -   | -   | +   | +   | -   | -   | -   |
| 49A | +   | -   | +   | +   | +   | -   | -   |
| GNA | +   | -   | +   | -   | +   | -   | +   |
| G39 | +   | -   | -   | -   | +   | -   | +   |

**Fig (4).** Cancer surveillance network using grid technology

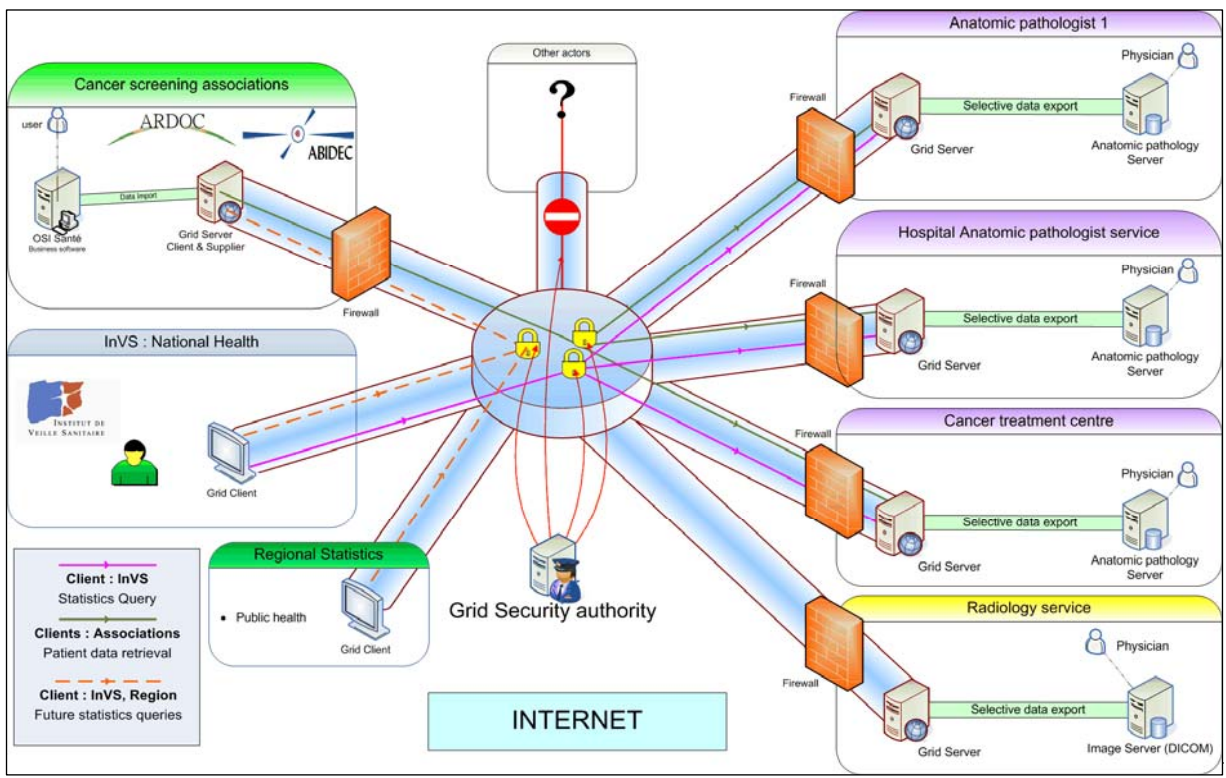

**GRID TECHNOLOGY ENABLED SURVEILLANCE NETWORK ON AVIAN FLU**

The ability of the international community to respond efficiently to the possible emergence of a human-to-human transmissible avian influenza virus depends on its capacity to quickly assess any evolution of the disease. The countries which have been most exposed to avian flu have set up very efficient national networks for collecting data and monitoring the outbreaks.

Unfortunately, there is no efficient international surveillance network so far that allows sharing the data collected at national level. However, more and more public resources are currently storing sequences of the virus strains and gathering useful information to the research community.

The Influenza Virus Resource (http://www.ncbi.nlm.nih.gov/genomes/FLU/FLU.html) at the National Center for Biotechnology Information (NCBI) and the Influenza Sequence Database (ISD) at Los Alamos National Laboratory are the two main repositories currently available. A significant effort has been made worldwide to sequence viral isolates in order to obtain nucleotide and protein information. Other resources focus more on curating and corroborating collected informations: this data processing is necessary to solve inaccuracy or repetition due to researchers direct submission to the database. Indeed the Influenza Virus Database (http://influenza.big.ac.cn/) from the Beijing Institute of Genomics - Chinese Academy of Sciences hosts around 50,000 sequences and provides tools to analyze genomes, genes, genetic polymorphisms and phylogenetic relationships. ISD also provide curated data but not for free **[22]**.

An interim Influenza Virus Tracking System (https://www.who.int/fluvirus_tracker) is hosted by the World Health Organization (WHO). The interim system contains data for the majority of viruses and clinical specimens that have been shared with this organization since November 2007, as well as all H5N1 viruses that have been developed into reassortant viruses as potential vaccine candidates. Data entry is ongoing. Through hundreds of missions, the Food and Agriculture Organization (FAO) has worked with affected and at-risk countries to facilitate capacity building, information sharing and networking. (http://www.fao.org/avianflu/en/index.html). FAO provides services to reinforce global avian influenza surveillance and early warning capabilities.

The proposed surveillance network would federate national data repositories on avian flu together with international public resources previously described. The services to implement in priority are those addressing the main concerns of the international community, namely the evolution of the virus and its capacity for human-to-human transmission. For this purpose, services for molecular epidemiology and *in silico* drug discovery are extremely relevant: they both rely on the virus sequences. Modelling of pandemics propagation appears also as an important service, which requires the availability of up to date geographical distribution of the outbursts.

The grid technology to build such a monitoring network has been developed in the recent years. Development of standards for interoperability allows a joint deployment across grid infrastructures all around the world.

Such a worldwide surveillance network would involve many stakeholders:

- International organizations like WHO or FAO.

- National public health institutes and centers for disease control like National Institute for Sanitary Watch

- Research laboratories on infectious diseases like Institut Pasteur.

These stakeholders act as data providers or customers of the services provided by the network.

More generally, the quality of the surveillance depends on several parameters:
- The design and deployment of a robust federation of databases on multiple sites worldwide.
- The reliability, relevance, completeness of the data stored. Reliability has to do with the mechanism to collect data. Relevance has to do with the mechanism to update data while completeness has to do with the capacity to collect data from multiple scientific disciplines and multiple countries.
- The relevance of the services offered. It is not sufficient to have the best, most up-to-date information on the disease. The data must be properly integrated and its exploitation must build upon expert skills in molecular epidemiology and bioinformatics.
- The user friendliness of the environment. Scientists are so busy they will not take time to contribute information and operate services if these services are not easy to use.
- The reliability of the security framework.

A key to the success is that users ought to be able to take for granted that the security mechanisms are sufficient to protect their data and that the results of their research will be private and available to third parties only if desirable. Indeed a significant delay remains - at least several months – before the new sequenced strains are included in the database to allow scientists to publish on these new sequences. This creates an unsatisfactory situation where possible mutations leading to human-to-human transmission could remain hidden for several weeks or months to the scientific community. Grid technology is now sufficiently mature to provide the secured framework fulfilling these requirements. Indeed, the grid certificates, created firstly to authenticate an user on the grid can easily be used to several purposes: set up a trusted client-server connexion between actors, perform rights managements between data holders and users or logging (connexion and usage of resources). Such a framework is currently being tested on the cancer surveillance network described in section 4 where the different anatomical pathology laboratories manage and control the usage of their data. In order to improve the security, the certificate can be stored on a personal smart card, widely use in France health system.

Applied to the avian flu surveillance network, it means that a biologist committing data to the network would keep the privilege of granting access rights to his/her data and possibly restricting it to collaborators or partners. In case of emergency, access to the data could be granted to an emergency panel for pandemics risk evaluation without compromising the ability for the biologist to later on publish analysis of the data.

We are going to further detail the services to be offered by the proposed surveillance network.

**Molecular epidemiology**

The H5N1 epidemic has been characterized by a constant evolution of the virus genetic content, as it is the case for all the other influenza viruses. The main risk of pandemics comes from mutations which would induce human-to-human transmission. Monitoring of the evolution of the different virus strains is therefore a necessity for the research community. It must include both human and bird cases and starts at a very local level whenever an outburst takes place. National networks have been set up to quickly collect virus sample, analyze it to confirm the origin of death and sequence the virus strain involved. Once the virus sequence is available, its phylogenetic analysis requires a comparison with all the other virus strains sequenced which is a CPU consuming task even for small viruses. A service for automatic computation of phylogenetic relationships between molecular sequences and reconstruction of a robust phylogenetic tree from a set of sequences is then performed on grid using the best open source phylogenetic software codes. A biological meaningful approach is the deployment of distributed workflows for comparative phyloinformatics using different algorithms for a constant monitoring of molecular evolution at both macro level (within the group of viruses) and micro level (within the group of strains) **[23]**.

Grid is a pleasant solution, with its large number of distributed CPU available, to deploy such workflows suffering from long runtimes due to their high degree of computational complexity as well as the large datasets involved.

**In silico drug discovery**

Once a new virus strain is identified, the 3D structure of its enzymes can be derived from the already available H5N1 structures in PDB using homology modelling. These new structures can be docked against the already known drugs to evaluate, *in silico*, the potential impact of the mutations on the drug efficiency.

Another approach - as explained in section 3 - is virtual screening on grid to identify possible new hits and/or new scaffolds. Grids enable storage and automatic update of databases of chemical compounds ready for molecular modelling allowing a rapid deployment of *in silico* experiments when a new crystallized or modelled 3D structure becomes available in research community. A similar initiative is already available through a global partnership forged over the PRAGMA grid development activities in order to build a scalable, global, and open knowledge environment for developing novel inhibitors: avian flu grid. (http://avianflugrid.pragma-grid.net/).

The use of *in silico* docking methods is sometimes the only possible strategy when the infectious agent cannot be propagated safely or with sufficient reproducibility in a laboratory environment. Results will provide important information concerning chemical features of potential inhibitors and their binding mode allowing further QSAR (Quantitative structure-activity relationship) studies.

**Pandemics propagation modelling**

If the virus acquires the ability to transmit from human-to-human, it will take more than 6 months to develop a vaccine to protect populations. Influenza pandemic planning is a complex, multifactorial process, raising the need to understand from now how influenza pandemics spread, both over time and geographically. In 2007, an international team of physicians and epidemiologists completed a study on the possible scenarios of the propagation of an avian influenza pandemic, under the hypothesis of person-to-person virus transmission **[24]**. Mathematical models allow to model temporal and geographical spread of avian flu based on virulence of transmission and conditions of its starting point (a flu starting in the summer does not propagate as easily as in winter).

At the laboratory MSI (IFI, Hanoi, Vietnam), in 2007-2008, a work on modelling of the spread of avian influenza in a province of North Vietnam in collaboration with CIRAD (French Agricultural Research Centre working for International Development) has been done. This is an application on a new simulation platform called GAMA (Gis & Agent-based Modelling Architecture) which aims at providing field experts, modellers, and computer scientists, a complete modelling and simulation development environment for building spatially explicit multi-agent simulations. The goal is to simulate the poultry value chain of a whole province using geo-localised data, and use this to optimize a monitoring network **[25]**. Moreover, the researchers claim that a combination of an agent-based model and a SIG-based environmental model can act as a "virtual laboratory" for epidemiology. Following the needs expressed by epidemiologists studying micro-scale dynamics of avian influenza in Vietnam, and after a review of the epidemiological models proposed so far, an improved model, always built on top of the GAMA platform has been developed and adapted to the epidemiologists' requirements. One notable contribution of this work is the treatment of the environment together with the social structure and animals' behaviours, as a first-class citizen in the model, allowing epidemiologists to consider heterogeneous micro and macro factors in their exploration of the causes of the epidemics **[26]**.

Grid technology can be a great solution to share and organize data in order to improve computational models and design probable scenarios. Grid can supply its calculation power to a larger simulation schema or allow even realizing a distributed simulation at world scale.

**CONCLUSION**

In this paper, we have discussed the opportunity for addressing evolution of a virus and its capacity for human-to-human transmission to deploy a surveillance network in the case of avian flu taking advantage of the recent developments of an emerging technology: the grid. We have shown how grids were already used to deploy innovative *in silico* approaches for molecular epidemiology and virtual screening on avian flu. From the field of oncology, we have provided the example of a surveillance network empowered by the grid where the different data sources are federated and queried remotely through a secured framework.

Many contacts have been taken with international organizations (FAO, WHO), research laboratories and Centers for Disease Control and Prevention throughout Africa (Egypt), America (US), Asia (China, Korea, Thailand, Vietnam, Taiwan), and Europe (France, Italy, UK) in the last year to foster the emergence of this network which is a huge enterprise and has to be done in a very progressive way. As a first step, the work will focus on setting up molecular epidemiology and virtual screening services on the EGEE grid infrastructure to demonstrate the relevance of the approach to the Public Health authorities.

**ACKNOWLEDGMENT**

The work described in this article was partly supported by grants from the European Commission (BioinfoGRID, EGEE, Embrace), the French Miinistry of Research (AGIR, GWENDIA) and the regional authorities (Conseil Régional d'Auvergne, Conseil Général du Puy-de-Dôme, Conseil Général de l'Allier).
The Enabling Grids for E-sciencE (EGEE) project is co-funded by the European Commission under contract INFSO-RI-031688. The BioinfoGRID project is co-funded by the European Commission under contract INFSO-RI-026808. The EMBRACE project is co-funded by the European Commission under the thematic area "Life sciences, genomics and biotechnology for health", contract number LHSG-CT-2004-512092. The SHARE project is co-funded by the European Commission under contract number FP6-2005-IST-027694.
Auvergrid is a project funded by the Conseil Regional d'Auvergne. The French ministry of Research supports the AGIR and GWENDIA projects.
*In vitro* work was partially supported by the Korean Foundation for International Cooperation of Science & Technology (KICOS) through a grant provided by the Korean Ministry of Science & Technology (MOST) in 2008 (No. K20812000001-08B1300-00110).
TWGrid funded by the National Science Council (NSC), Taiwan participated partly in deploy DIANE for virtual screening on avian flu and post data analysis.